\documentclass[letter paper,12pt]{article}
\usepackage{float}
\usepackage{natbib}
\usepackage{amsmath}
\usepackage{amsfonts}
\usepackage{graphicx}
\usepackage{epstopdf}
\usepackage[FIGTOPCAP,tight]{subfigure}

\newcommand{\bm}[1]{\mbox{\protect\boldmath $ #1 $}}
\newtheorem{definition}{Definition}[section]    
\newtheorem{theorem}[definition]{Theorem}

\floatstyle{plain}
\newfloat{algorithm}{htbp}{loa}
\floatname{algorithm}{Algorithm}

\newcommand{\EE}[1]{\mathbb{E} \left( #1 \right)}

\usepackage{titlesec}
\titleformat{\section}{%
\centering\normalfont\normalsize}{\thesection.}{1em}{}
\titleformat{\subsection}{%
\normalfont\normalsize}{\thesubsection}{1em}{}

\graphicspath{{./Plots/}}

\title{Optimal detection of changepoints with a linear computational cost}
\author{Killick, R., Fearnhead, P. and Eckley, I.A.\footnote{R. Killick is Senior Research Associate, Department of Mathematics \& Statistics, Lancaster University, Lancaster, UK (E-mail:
r.killick@lancs.ac.uk). P. Fearnhead is Professor, Department of Mathematics \& Statistics, Lancaster University, Lancaster, UK (E-mail: p.fearnhead@lancs.ac.uk). I.A. Eckley is Senior Lecturer,
Department of Mathematics \& Statistics, Lancaster University, Lancaster, UK (E-mail: i.eckley@lancs.ac.uk). The authors are grateful to Richard Davis and Alice Cleynen for providing the Auto-PARM and
PDPA software respectively.  Part of this research was conducted whilst R. Killick was a jointly funded Engineering
and Physical Sciences Research Council (EPSRC) / Shell Research Ltd graduate student at Lancaster University. Both I.A. Eckley and R. Killick also gratefully acknowledge the financial support of the
EPSRC grant number EP/I016368/1.
}}

\begin{document}
\maketitle

\begin{abstract}
We consider the problem of detecting multiple changepoints in large data sets. Our focus is on applications where the number of changepoints will increase as we collect more data: 
for example in genetics as we analyse larger regions of the genome, or in finance as we observe time-series over longer periods. We consider the common approach of detecting changepoints through
minimising a cost function over possible numbers and locations of changepoints. This includes several established procedures for detecting changing points, such as penalised likelihood and minimum
description length. We introduce a new method for finding the minimum of such cost functions and hence the optimal number and location of changepoints that has a computational cost which, under mild
conditions, is linear in the number of observations. This compares favourably with existing methods for the same problem whose computational cost can be quadratic or even cubic. In simulation studies
we show that our new method can be orders of magnitude faster than these alternative exact methods. We also compare with the Binary Segmentation algorithm for identifying changepoints, showing that
the exactness of our approach can lead to substantial improvements in the accuracy of the inferred segmentation of the data. 
\end{abstract}

\section{INTRODUCTION}
As increasingly longer data sets are being collected, more and more applications require the detection of changes in the distributional properties of such data.
Consider for example recent work in genomics, looking at detecting changes in gene copy numbers or in the compositional structure of the genome \cite[]{BraunBraunMuller2000,Olshen:2004,PicardRobinLavielleVaisseDaudin2005}; 
 and in finance where, for example, interest lies in detecting changes in the volatility of time series \cite[]{AggarwalInclanLeal1999,AndreouGhysels2002,Fernandez2004}. 
 Typically such series will contain several changepoints.  There is therefore a growing need to be able to search for such changes efficiently.  
It is this search problem which we consider in this paper. In particular we focus on applications where we expect the number of changepoints to increase as we collect more data. 
This is a natural assumption in many cases, for example as we analyse longer regions of the genome or as we record financial time-series over longer time-periods. By comparison it does not necessarily apply to situations where we are obtaining data over a fixed time-period at a higher frequency.

At the time of writing Binary Segmentation proposed by \cite{ScottKnott1974} is arguably the most widely used changepoint search method. It is approximate in nature with an $\mathcal{O}(n\log{n})$
computational cost, where $n$ is the number of data points. While exact search algorithms exist for the most common forms of changepoint models, these have a much greater computational cost. Several
exact search methods are based on dynamic programming.  For example the Segment Neighbourhood method proposed by \cite{AugerLawrence1989} is $\mathcal{O}(Qn^2)$, where $Q$ is the maximum number of
changepoints you wish to search for. Note that in scenarios where the number of changepoints increases linearly with $n$, this can correspond to a computational cost that is cubic in the length of the
data.  An alternative dynamic programming algorithm is provided by the Optimal Partitioning approach of \cite{Jacksonetal2005}.  As we describe in Section \ref{Sec:alg} this can be applied to a
slightly smaller class of 
problems and is an exact approach whose computational cost is $\mathcal{O}(n^2)$. 

We present a new approach to search for changepoints, which is exact and under mild conditions has a computational cost that is linear in the number of data points: 
the {\bf P}runed {\bf E}xact {\bf L}inear {\bf T}ime (PELT) method. This approach is based on the algorithm of \cite{Jacksonetal2005}, but involves a pruning step within 
the dynamic program. This pruning reduces the computational cost of the method, but does not affect the exactness of the resulting segmentation. 
It can be applied to find changepoints under a range of statistical criteria such as penalised likelihood, quasi-likelihood 
\cite[]{BraunBraunMuller2000} and cumulative sum of squares \cite[]{InclanTiao1994,PicardLebarbierHoebekeRigaillThiamRobin2011}.
In simulations we compare PELT with both Binary Segmentation and Optimal Partitioning. We show that PELT can be calculated orders of magnitude faster than Optimal Partitioning, particularly for long data sets. 
Whilst asymptotically PELT can be quicker, we find that in practice Binary Segmentation is quicker on the examples we consider, and we believe this would be the case in almost all applications. 
However, we show that PELT leads to a substantially more accurate segmentation than Binary Segmentation. 

The paper is organised as follows.  We begin in Section \ref{Sec:cptbgd} by reviewing some basic changepoint notation and summarizing existing work in the area of search methods. 
 The PELT method is introduced in Section \ref{Sec:algO} and the computational cost of this approach is considered in Section \ref{Sec:linearcost}. 
 The efficiency and accuracy of the PELT method are demonstrated in Section \ref{Sec:simstudy}.  In particular we demonstrate the methods' performance on large data sets coming from oceanographic (Section \ref{sec:NS}) and financial (supplementary material) applications. Results show the speed gains over other exact search methods and the increased accuracy relative to approximate search methods such as Binary Segmentation.  The paper concludes with a discussion.

\section{BACKGROUND}\label{Sec:cptbgd}
Changepoint analysis can, loosely speaking, be considered to be the identification of points within a data set where the statistical properties change.  More formally, let us assume we have an ordered sequence of data, ${y}_{1:n}=(y_1,\ldots,y_n)$.  Our model will have a number of changepoints, $m$, together with their positions, $\tau_{1:m}=(\tau_1,\ldots,\tau_m)$. Each changepoint position is an integer between 1 and $n-1$ inclusive. We define $\tau_0=0$ and $\tau_{m+1}=n$ and assume that the changepoints are ordered such that $\tau_i<\tau_j$ if, and only if, $i<j$.  Consequently the $m$ changepoints will split the data into $m+1$ segments, with the $i$th segment containing $y_{(\tau_{i-1}+1):\tau_{i}}$.

One commonly used approach to identify multiple changepoints is to minimise:
\begin{align}
	\sum_{i=1}^{m+1}{\left[\mathcal{C}(y_{(\tau_{i-1}+1):\tau_i})\right] + \beta f(m)}. \label{eqn:aim}
\end{align}
Here $\mathcal{C}$ is a cost function for a segment and $\beta f(m)$ is a penalty to guard against over fitting.  
Twice the negative log likelihood is a commonly used cost function in the changepoint literature \cite[see for example][]{Horvath1993,ChenGupta2000}, 
although other cost functions such as quadratic loss and cumulative sums are also used  \cite[e.g.\ ][]{Rigaill2010,InclanTiao1994}, or those based on both the segment log-likelihood and the length of
the segment \cite[]{ZhangSiegmund2007}.
 Turning to choice of penalty, in practice by far the most common choice is one which is linear in the number of changepoints, i.e.\ $\beta f(m) = \beta m$.  
Examples of such penalties include Akaike's Information Criterion (AIC, \cite{Akaike1974})  $(\beta=2p)$ and Schwarz Information Criterion \cite[SIC, also known as BIC;][]{Schwarz1978} $(\beta=p\log{n})$, where $p$ is the number of additional parameters introduced by adding a changepoint. 
The PELT method which we introduce in Section \ref{Sec:algO} is designed for such linear cost functions.  
Although linear cost functions are commonplace within the changepoint literature \cite{GuyonYao1999}, \cite{PicardRobinLavielleVaisseDaudin2005} and \cite{BirgeMassart2007} offer examples and discussion of alternative penalty choices. 
In Section \ref{Sec:conca} we show how PELT can be applied to some of these alternative choices. 

The remainder of this section describes two commonly used methods for multiple changepoint detection; Binary Segmentation \citep{ScottKnott1974} and Segment Neighbourhoods \citep{AugerLawrence1989}.  A third method proposed by \cite{Jacksonetal2005} is also described as it forms the basis for the PELT method which we propose.  For notational simplicity we describe all the algorithms (including PELT) assuming that the minimum segment length is a single observation, i.e. $\tau_{i-1}-\tau_i\geq 1$. A larger minimum segment length is easily implemented when appropriate, see for example Section \ref{Sec:simstudy}.

\subsection{Binary Segmentation}
Binary Segmentation (BS) is arguably the most established search method used within the changepoint literature.  Early applications include \cite{ScottKnott1974} and \cite{SenSrivastava1975}.  In essence the method extends any single changepoint method to multiple changepoints by iteratively repeating the method on different subsets of the sequence.  It begins by initially applying the single changepoint method to the entire data set, i.e.\ we test if a $\tau$ exists that satisfies
\begin{align}
	\mathcal{C}(y_{1:\tau}) + \mathcal{C}(y_{(\tau+1):n}) +\beta < \mathcal{C}(y_{1:n}). \label{eqn:binseg1}
\end{align}
If \eqref{eqn:binseg1} is false then no changepoint is detected and the method stops.  Otherwise the data is split into two segments consisting of the sequence before and after the identified changepoint, $\tau_{a}$ say, 
and apply the detection method to each new segment.  
If either or both tests are true, we split these into further segments at the newly identified changepoint(s), applying the detection method to each new segment. This procedure is repeated until no further changepoints are detected.  For pseudo-code of the BS method see for example \cite{EckleyFearnheadKillick2010}.

Binary Segmentation can be viewed as attempting to minimise equation \eqref{eqn:aim} with $f(m)=m$: each step of the algorithm attempts to introduce an extra changepoint if and only if it reduces \eqref{eqn:aim}.
The advantage of the BS method is that it is computationally efficient, resulting in an $\mathcal{O}(n\log{n})$ calculation.  
However this 
comes at a cost as it is not guaranteed to find the global minimum of \eqref{eqn:aim}.

\subsection{Exact methods}\label{Sec:alg}
\paragraph{{\itshape Segment Neighbourhood}}
\cite{AugerLawrence1989} propose an alternative, exact search method for changepoint detection, namely the Segment Neighbourhood (SN) method.  This approach searches the entire segmentation space
using dynamic programming \citep{BellmanDreyfus1962}. It begins by setting an upper limit on the size of the segmentation space (i.e.\ the maximum number of changepoints) that is required -- this is
denoted $Q$.  The method then continues by computing the cost function for all possible segments.  From this all possible segmentations with between 0 and $Q$ changepoints are considered.  

In addition to being an exact search method, the SN approach has the ability to incorporate an arbitrary penalty of the form, $\beta f(m)$.  However, a consequence of the exhaustive search is that the method has significant computational cost, $\mathcal{O}(Qn^2)$.  If as the observed data increases, the number of changepoints increases linearly, then $Q=\mathcal{O}(n)$ and the method will have a computational cost of $\mathcal{O}(n^3)$. 

\paragraph{{\itshape The optimal partitioning method}}
\cite{Yao1984} and \cite{Jacksonetal2005} propose a search method that aims to minimise
\begin{align}
	\sum_{i=1}^{m+1}{\left[\mathcal{C}(y_{(\tau_{i-1}+1):\tau_i}) + \beta \right]}. \label{eqn:aimlinear}
\end{align}
This is equivalent to (\ref{eqn:aim}) where $f(m)=m$.

Following \cite{Jacksonetal2005} the optimal partitioning (OP) method begins by first conditioning on the last point of change.  It then relates the optimal value of the cost function to the cost for the optimal partition of the data prior to the last changepoint plus the cost for the segment from the last changepoint to the end of the data.
More formally, let $F(s)$ denote the minimisation from \eqref{eqn:aimlinear} for data $y_{1:s}$ and $\mathcal{T}_s=\{\bm{\tau}: 0=\tau_0<\tau_1<\cdots<\tau_{m}<\tau_{m+1}=s \}$ 
be the set of possible vectors of changepoints for such data. Finally set $F(0)=-\beta$. It therefore follows that:
\begin{align*}
	F(s) &= \min_{\bm{\tau}\in\mathcal{T}_s} \left\{\sum_{i=1}^{m+1}{\left[\mathcal{C}(y_{(\tau_{i-1}+1):\tau_i}) + \beta \right]}\right\},  \\
	 &=\min_{t}\left\{\min_{\bm{\tau}\in\mathcal{T}_{t}} \sum_{i=1}^m \left[\mathcal{C}(y_{(\tau_{i-1}+1):\tau_i})+\beta\right]+\mathcal{C}(y_{(t+1):n})+\beta \right\}, \\
&=\min_{t}\left\{F(t)+\mathcal{C}(y_{(t+1):n})+\beta \right\}.
\end{align*}
This provides a recursion which gives the minimal cost for data $y_{1:s}$ in terms of the minimal cost for data $y_{1:t}$ for $t<s$. This recursion can be solved in turn for $s=1,2,\ldots,n$. The cost of solving the recursion for time $s$ 
is linear in $s$, so the overall computational cost of finding $F(n)$ is quadratic in $n$. Steps for implementing the OP method are given in Algorithm \ref{alg:KFE}.

\begin{algorithm}[t]
\begin{footnotesize}
\rule{\textwidth}{1mm}
\underline{Optimal Partitioning}

\begin{tabular}[h]{ll}
{\bf Input:} & A set of data of the form, $(y_1,y_2,\ldots,y_n)$ where $y_i \in \mathbb{R}$. \\
& A measure of fit $\mathcal{C}(\cdot)$ dependent on the data. \\
& A penalty constant $\beta$ which does not depend on the number or location of changepoints. \\
{\bf Initialise:} & Let $n=$ length of data and set $F(0)=-\beta$, $cp(0)=NULL$.
\end{tabular}
{\bf Iterate} for $\tau^*=1,\ldots,n$
\begin{enumerate}
\item Calculate $F(\tau^*) = \min_{0\leq\tau < \tau^*}\left[ F(\tau)+\mathcal{C}(y_{(\tau+1):\tau^*})+ \beta \right]$.
\item Let $\tau'=\arg \left\{\min_{0\leq\tau < \tau^*}\left[ F(\tau)+\mathcal{C}(y_{(\tau+1):\tau^*})+ \beta \right]\right\}$.
\item Set $cp(\tau^*) = (cp(\tau'),\tau')$.
\end{enumerate}

{\bf Output} the change points recorded in $cp(n)$.

\rule{\textwidth}{1mm}
\caption{{\footnotesize Optimal Partitioning.}}\label{alg:KFE}
\end{footnotesize}
\end{algorithm}

Whilst OP improves on the computational efficiency of the SN method, it is still far from being competitive computationally with the BS method. 
Section \ref{Sec:algO} introduces a modification of the optimal partitioning method denoted PELT which results in an approach whose computational cost can be linear in $n$ whilst retaining an exact
minimisation of \eqref{eqn:aimlinear}. 
This exact and efficient computation is achieved via a combination of optimal partitioning and pruning. 

\section{A PRUNED EXACT LINEAR TIME METHOD}\label{Sec:algO}
We now consider how pruning can be used to increase the computational efficiency of the OP method whilst still ensuring that the method finds a global minimum of the cost function \eqref{eqn:aimlinear}. 
The essence of pruning in this context is to remove those values of $\tau$ which can never be minima from the minimisation performed at each iteration in (1) of Algorithm \ref{alg:KFE}. 

The following theorem gives a simple condition under which we can do such pruning. 
\begin{theorem}\label{thm:condition}
We assume that when introducing a changepoint into a sequence of observations the cost, $\mathcal{C}$, of the sequence reduces.  More formally, we assume there exists a constant $K$ such that for all $t<s<T$, 
\begin{align}
	\mathcal{C}(y_{(t+1):s}) + \mathcal{C}(y_{(s+1):T}) +K \leq \mathcal{C}(y_{(t+1):T}).\label{ass:cpt}
\end{align}
Then if
\begin{align}
	F(t) + \mathcal{C}(y_{(t+1):s}) +K \geq F(s) \label{eqn:prune}
\end{align}
holds, at a future time $T>s$, $t$ can never be the optimal last changepoint prior to $T$.
\end{theorem}
{\bf Proof.} See Section 5 of Supplementary Material. \hfill $\Box$

The intuition behind this result is that if \eqref{eqn:prune} holds then for any $T>s$ the best segmentation with the most recent changepoint prior to $T$ being at $s$ will be better than any which has this most recent changepoint at $t$. Note that almost all cost functions used in practice satisfy assumption \eqref{ass:cpt}. For example, if we take the cost function to be minus the log-likelihood then the constant $K=0$ and if we take it to be minus a penalised log-likelihood then $K$ would equal the penalisation factor.

The condition imposed in Theorem \ref{thm:condition} for a candidate changepoint, $t$, to be discarded from future consideration is important as it removes computations that are not relevant for obtaining the final set of changepoints. This condition can be easily implemented into the OP method and the pseudo-code is given in Algorithm \ref{alg:KFEp}.  This shows that at each step in the method the candidate changepoints satisfying the condition are noted and removed from the next iteration. We show in the next section that under certain conditions the computational cost of this method will be linear in the number of observations, as a result we call this the {\bf P}runed {\bf E}xact {\bf L}inear {\bf T}ime (PELT) method.

\begin{algorithm}[t]
\begin{footnotesize}
\rule{\textwidth}{1mm}
\underline{PELT Method}

\begin{tabular}[h]{ll}
{\bf Input:} & A set of data of the form, $(y_1,y_2,\ldots,y_n)$ where $y_i \in \mathbb{R}$. \\
& A measure of fit $\mathcal{C}(.)$ dependent on the data. \\
& A penalty constant $\beta$ which does not depend on the number or location of changepoints. \\
& A constant $K$ that satisfies equation \ref{ass:cpt}.\\
{\bf Initialise:} & Let $n=$ length of data and set $F(0)=-\beta$, $cp(0)=NULL$, $R_1=\{0\}$. 
\end{tabular}
{\bf Iterate} for $\tau^*=1,\ldots,n$
\begin{enumerate}
\item Calculate $F(\tau^*) = \min_{\tau \in R_{\tau^*}}\left[ F(\tau)+\mathcal{C}(y_{(\tau+1):\tau^*})+ \beta \right]$.
\item Let $\tau^1=\arg \left\{\min_{\tau \in R_{\tau^*}}\left[ F(\tau)+\mathcal{C}(y_{(\tau+1):\tau^*})+ \beta \right]\right\}$.
\item Set $cp(\tau^*) = [cp(\tau^1),\tau^1]$.
\item Set $R_{\tau^*+1}=\left\{\tau\in R_{\tau^*} \cup \{\tau^*\}: F(\tau)+\mathcal{C}(y_{\tau+1:\tau^*}) +K  \leq F(\tau^*) \right\}$.
\end{enumerate}

{\bf Output} the change points recorded in $cp(n)$.

\rule{\textwidth}{1mm}
\caption{{\footnotesize PELT Method.}}\label{alg:KFEp}
\end{footnotesize}
\end{algorithm}

\subsection{Linear Computational Cost of PELT}\label{Sec:linearcost}
We now investigate the theoretical computational cost of the PELT method. We focus on the most important class of changepoint models and penalties and provide sufficient conditions for the method to have a computational cost 
that is linear in the number of data points. 
 The case we focus on is the set of models where the 
segment parameters are independent across segments and the cost function for a segment is minus the maximum log-likelihood value for the data in that segment.

More formally, our result relates to the expected computational cost of the method and how this depends on the number of data points we analyse. To this end we define an underlying stochastic model for the data generating process. Specifically we define such a process over positive-integer time points and then consider analysing the first $n$ data points generated by this process. Our result assumes that the parameters associated with a given segment are IID with density function $\pi(\theta)$.  For notational simplicity we assume that given the parameter, $\theta$, for a segment, the data points within the segment are IID with density function $f(y|\theta)$ although extensions to dependence within a segment is trivial. Finally, as previously stated our cost function will be based on minus the maximum log-likelihood:
\[
\mathcal{C}(y_{(t+1):s})= - \max_\theta \sum_{i=t+1}^s \log f(y_i|\theta).
\]
Note that for this loss-function, $K=0$ in \eqref{ass:cpt}.  Hence pruning in PELT will just depend on the choice of penalty constant $\beta$.

We also require a stochastic model for the location of the changepoints in the form of a model for the length of each segment. If the changepoint positions are $\tau_1,\tau_2,\ldots,$ then define the segment lengths to be $S_1=\tau_1$ and for $i=2,3,\ldots,$ $S_i=\tau_{i}-\tau_{i-1}$. We assume the $S_i$ are IID copies of a random variable $S$. Furthermore $S_1,S_2,\ldots,$ are independent of the parameters associated with the segments. 

\begin{theorem} \label{thm:linear}
Define $\theta^*$ to be the value that maximises the expected log-likelihood
\[
\theta^*=\arg\max \int\int  f(y|\theta) f(y|\theta_0) \mbox{d}y \pi(\theta_0) \mbox{d}\theta_0.
\]
Let $\theta_i$ be the true parameter associated with the segment containing $y_i$ and $\hat\theta_n$ be the maximum likelihood estimate for $\theta$ given data $y_{1:n}$ and an assumption of a single segment:
\[
\hat\theta_n=\arg\max_\theta \sum_{i=1}^n \log f(y_i|\theta).
\]
Then if
\begin{itemize}
\item[(A1)] denoting
\[
B_n=\sum_{i=1}^n \left[\log f(y_i|\hat\theta_n)-\log f(y_i|\theta^*)\right],
\]
we have $\EE{B_n}=o(n)$ and $\EE{[B_n-\EE{B_n}]^4}=\mathcal{O}(n^2)$;
\item[(A2)] \[
\EE{ \left[ \log f(Y_i|\theta_i) - \log f(Y_i|\theta^*) \right]^4}<\infty;
\]
\item[(A3)] \[\EE{S^4}<\infty;\mbox{ and}\]
\item[(A4)] \[
\EE{ \log f(Y_i|\theta_i) - \log f(Y_i|\theta^*)} > \frac{\beta}{\EE{S}};
\]
\end{itemize}
where $S$ is the expected segment length, the expected CPU cost of PELT for analysing $n$ data points is bounded above by $Ln$ for some constant $L<\infty$.
\end{theorem}
{\bf Proof.} See Section 6 of Supplementary Material. \hfill $\Box$

Conditions (A1) and (A2) of Theorem \ref{thm:linear} are weak technical conditions. For example, general asymptotic results for maximum likelihood estimation would give $B_n=\mathcal{O}_p(1)$, and (A1) is a slightly stronger condition which is controlling the probability of $B_n$ taking values that are $\mathcal{O}(n^{1/2})$ or greater. 

The other two conditions are more important. Condition (A3) is needed to control the probability of large segments. One important consequence of (A3) is that the expected number of changepoints will increase linearly with $n$. Finally condition (A4) is a natural one as it is required for the expected penalised likelihood value obtained with the true changepoint and parameter values to be greater than the expected penalised likelihood value if we fit a single segment to the data with segment parameter $\theta^*$. 

In all cases the worst case complexity of the algorithm is where no pruning occurs and the computational cost is $\mathcal{O}(n^2)$.

\subsection{PELT for concave penalties} \label{Sec:conca}

There is a growing body of research  \cite[see][]{GuyonYao1999,PicardRobinLavielleVaisseDaudin2005, BirgeMassart2007} that consider nonlinear penalty forms.  
In this section we address how PELT can be applied to penalty functions which are concave.
\begin{align} \label{eq:MDL}
 \beta f(m) + \sum_{i=1}^{m+1}{\mathcal{C}(y_{(\tau_{i-1}+1):\tau_i})},
\end{align}
where $f(m)$ is concave and differentiable.

For an appropriately chosen $\gamma$, the following result shows that the optimum segmentation based on such a penalty corresponds to minimising 
\begin{equation} \label{eq:MDLrel}
 m \gamma + \sum_{i=1}^{m+1}{\mathcal{C}(y_{(\tau_{i-1}+1):\tau_i})}.
\end{equation}

\begin{theorem}
Assume that $f$ is concave and differentiable, with derivative denoted $f'$.  Further, let $\hat{m}$ be the value of $m$ for which the criteria (\ref{eq:MDL}) is minimised. Then the optimal segmentation under this set of penalties is the segmentation that minimises
\begin{equation} \label{eq:MDLp}
 m f'(\hat{m})+ \sum_{i=1}^{m+1}{\mathcal{C}(y_{(\tau_{i-1}+1):\tau_i})}.
\end{equation}
\end{theorem}
{\bf Proof.} See Section 7 of Supplementary Material. \hfill $\Box$

This suggests that we can minimize penalty functions based on $f(m)$ using PELT -- the correct penalty constant just needs to be applied. 
A simple approach, is to run PELT with an arbitrary penalty constant, say $\gamma=f'(1)$. Let $m_0$ denote the resulting number of changepoints estimated. 
We then run PELT with penalty constant $\gamma=f'(m_0)$, and get a new estimate of the number of changepoints $m_1$. If $m_0=m_1$ we stop.  Otherwise we update the penalty constant and repeat until convergence. 
This simple procedure is not guaranteed to find the optimal number of changepoints.  Indeed more elaborate search schemes may be better.  However, as tests of this simple approach in Section
\ref{sec:AR} show, it can be quite effective.

\section{SIMULATION AND DATA EXAMPLES}\label{Sec:simstudy}

We now compare PELT with both Optimal Partitioning (OP) and Binary Segmentation (BS) on a range of simulated and real examples. Our aim is to see empirically (i) how the computational cost of PELT is affected by the amount of data, (ii) to evaluate the computational savings that PELT gives over OP, and (iii) to evaluate the increased accuracy of exact methods over BS. Unless otherwise stated, we used the SIC penalty. In this case the penalty constant increases with the amount of data, and as such the application of PELT lies outside the conditions of Theorem \ref{thm:linear}. We also consider the impact of the number of changepoints not increasing linearly with the amount of data, a further violation of the conditions of Theorem \ref{thm:linear}.

\subsection{Changes in Variance within Normally Distributed Data}\label{Sec:changevar}\label{Sec:varsimstudy}
In the following subsections we consider multiple changes in variance within data sets that are assumed to follow a Normal distribution with a constant (unknown) mean.  
We begin by showing the power of the PELT method in detecting multiple changes via a simulation study, and then use PELT to analyse Oceanographic data and Dow Jones Index returns (Section
2 in supplementary material).

\paragraph{{\itshape Simulation Study}}
In order to evaluate PELT  we shall construct sets of simulated data on which we shall run various multiple changepoint methods.  It is reasonable to set the cost function, $\mathcal{C}$ as twice the
negative log-likelihood. 
 Note that for a change in variance (with unknown mean), the minimum segment length is two observations.  The cost of a segment is then
\begin{align}
 	\mathcal{C}(y_{(\tau_{i-1}+1):\tau_i}) = (\tau_i-\tau_{i-1})\left(\log(2\pi) + \log\left(\frac{\sum_{j=\tau_{i-1}+1}^{\tau_i}{(y_j - \mu)^2}}{\tau_i - \tau_{i-1}}\right)+1\right).
\end{align}

\begin{figure}[t]
	\caption{{\footnotesize A realisation of multiple changes in variance where the true changepoint locations are shown by vertical lines.}}\label{fig:varex}
	\begin{center}
		\includegraphics[width=\textwidth]{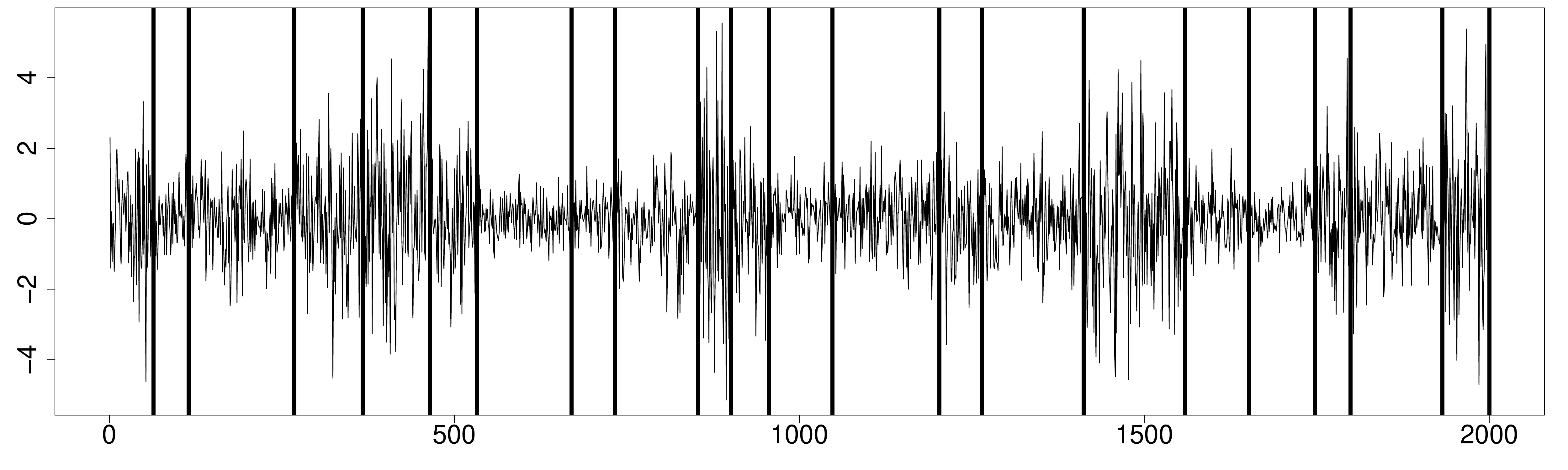}
	\end{center}
\end{figure}

Our simulated data consists of scenarios with varying lengths, $n=$(100, 200, 500, 1000, 2000, 5000, 10000, 20000, 50000). For each value of $n$ we
consider 
a linearly increasing number of changepoints, $m=n/100$.
In each case the changepoints are distributed uniformly across $(2,n-2)$ with the only constraint being that there must be at least 30 observations between changepoints.  
Within each of these scenarios we have 1,000 repetitions where the mean is fixed at 0 and the variance parameters for each segment are assumed to have a Log-Normal distribution with mean 0 
and standard deviation $\tfrac{\log(10)}{2}$. These parameters are chosen so that 95\% of the simulated variances are within the range $\left[\tfrac{1}{10},10\right]$. 
 An example realisation is shown in Figure \ref{fig:varex}. 
 Additional simulations considering a wider range of options for the number of changepoints (square root: $m=\lfloor\sqrt{n}/4\rfloor$ and fixed: $m=2$) and parameter values are given in Section 1 of
the Supplementary Material.

Results are shown in Figure \ref{fig:var} where we denote the Binary Segmentation method which identifies the \emph{same} number of changepoints as PELT as subBS. Conversely the number of changepoints
BS would optimally select is called optimal BS.  Firstly Figure \ref{fig:arlvar} shows that when the number of changepoints increases linearly with $n$, 
PELT does indeed have a CPU cost that is linear in $n$. By comparison figures in the supplementary material show that if the number of changepoints increases at a slower rate, for example, 
square root or even fixed number of changepoints, the CPU cost of PELT is no longer linear. However even in the latter two cases, substantial computational savings are attained relative to OP.
Comparison of times with BS are also given in the Supplementary material.  These show that PELT and BS have similar computational costs for the case of linearly increasing number of changepoints, but BS can be orders of magnitude quicker for other situations.

\begin{figure}[t]
	\caption{{\footnotesize (a) Average Computational Time (in seconds) for a change in variance (thin: OP, thick: PELT). (b) Average difference in cost between PELT and BS for subBS (thin), optimal BS
(thick)) (c) MSE for PELT (thick), optimal BS (thin) and subBS (dotted).
}}\label{fig:var}
	\centering
		\subfigure[]{\includegraphics[width=0.32\textwidth]{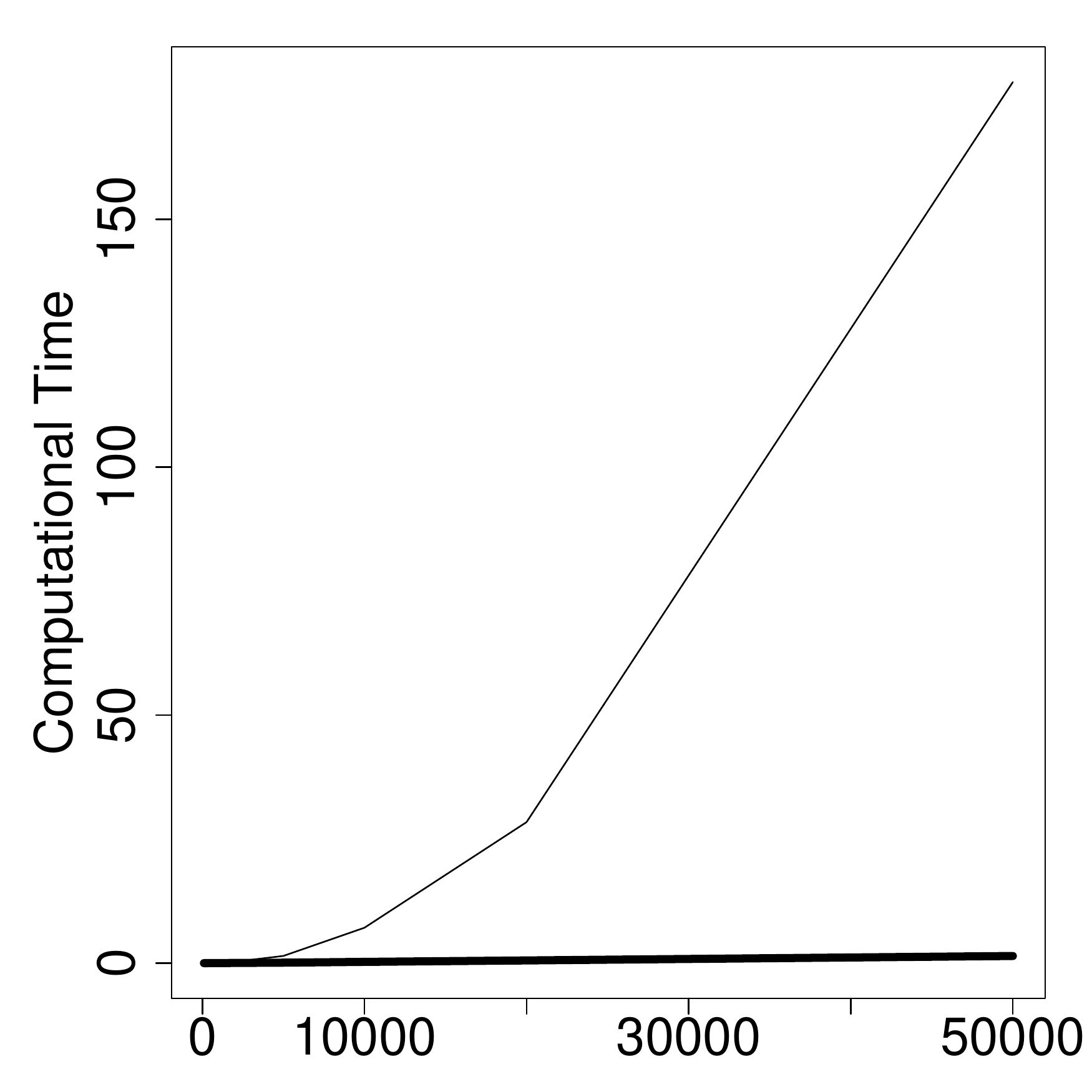}\label{fig:arlvar}}
		\subfigure[]{\includegraphics[width=0.32\textwidth]{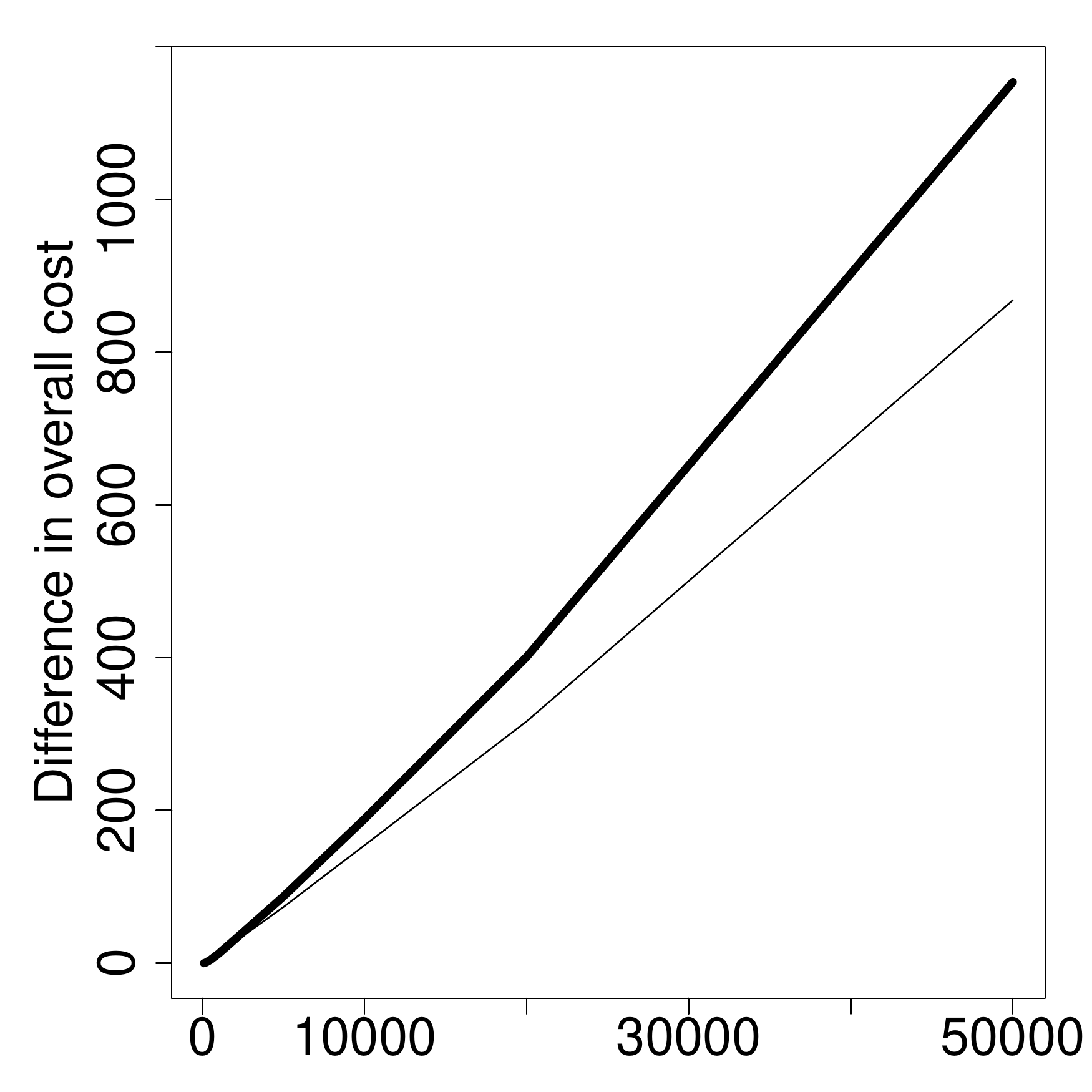}\label{fig:dcostvar}}
		\subfigure[]{\includegraphics[width=0.32\textwidth]{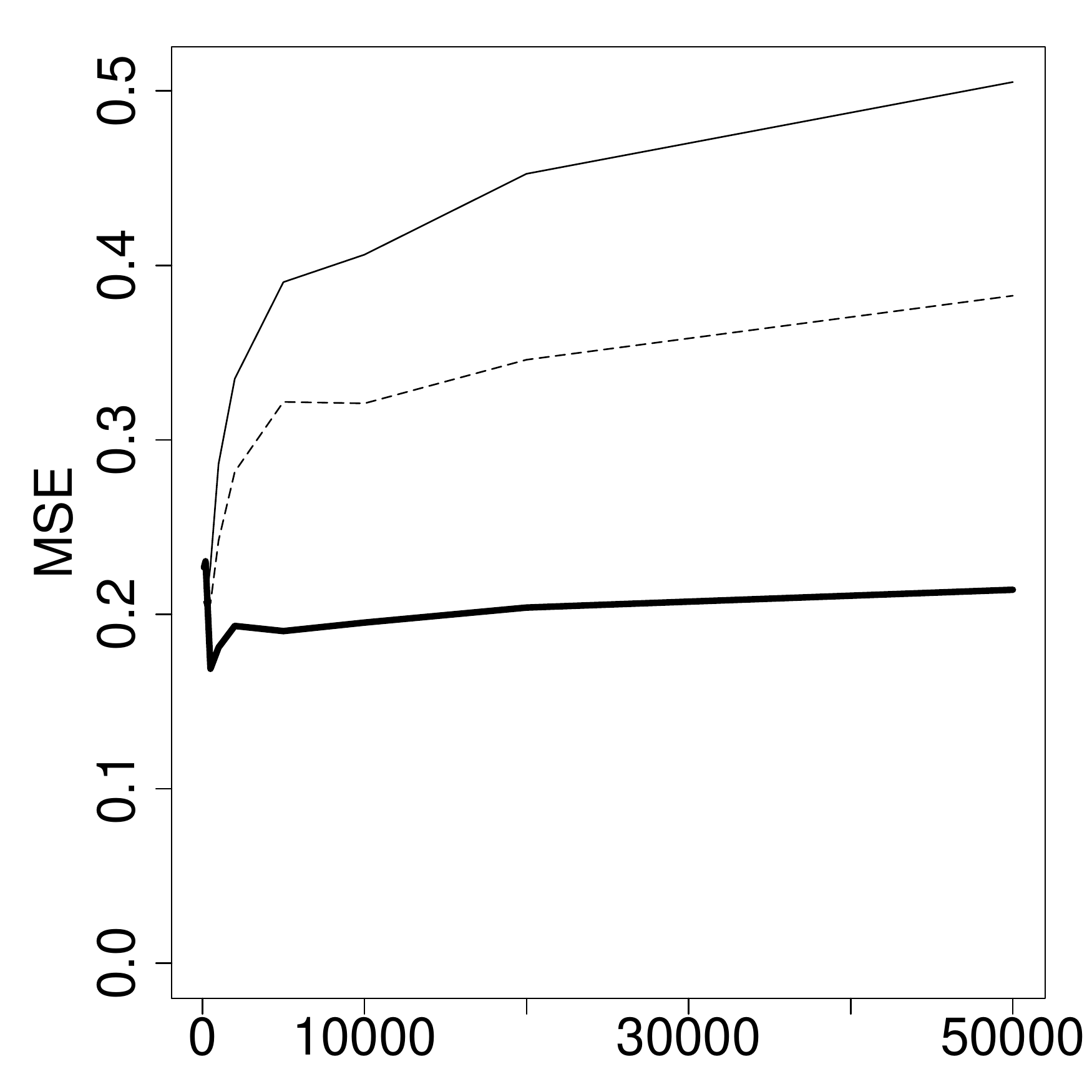}\label{fig:MSEvar}}
\end{figure}

The advantage of PELT over BS is that PELT is guaranteed to find the optimal segmentation under the chosen cost function, and as such is likely to be preferred providing sufficient computational time is available to run it. 
Figure \ref{fig:dcostvar} shows the improved fit to the data that PELT attains over BS in terms of the smaller values of the cost function that are found.  If you consider using the log-likelihood to choose between competing models, the value for $n=50,000$ is over 1000 which is very large.
An alternative comparison is to look at how well each method estimates the parameters in the model. We measure this using mean square error:
\begin{align}
\mbox{MSE } = \frac{\sum_{i=1}^n{(\hat{\theta}_i-\theta_i})^2}{n}, \label{eqn:MSE}
\end{align}
Figure \ref{fig:MSEvar} shows the increase in accuracy in terms of mean square error of estimates of the parameter. The figures in the supplementary material show that for the fixed number of changepoints scenario the difference is negligible but, for the linearly increasing number of changepoints scenario, the difference is relatively large. 

\begin{figure}[t]
	\caption{{\footnotesize Proportion of correctly identified changepoints against the proportion of falsely detected changepoints.  Change in variance with $m=n/100$ where (a) $n=500$, (b) $n=5,000$, (c) $n=50,000$ (PELT: thick line, BS: thin line, +: SIC penalty).}}\label{fig:FTvar}
	\centering
		\subfigure[]{\includegraphics[width=0.32\textwidth]{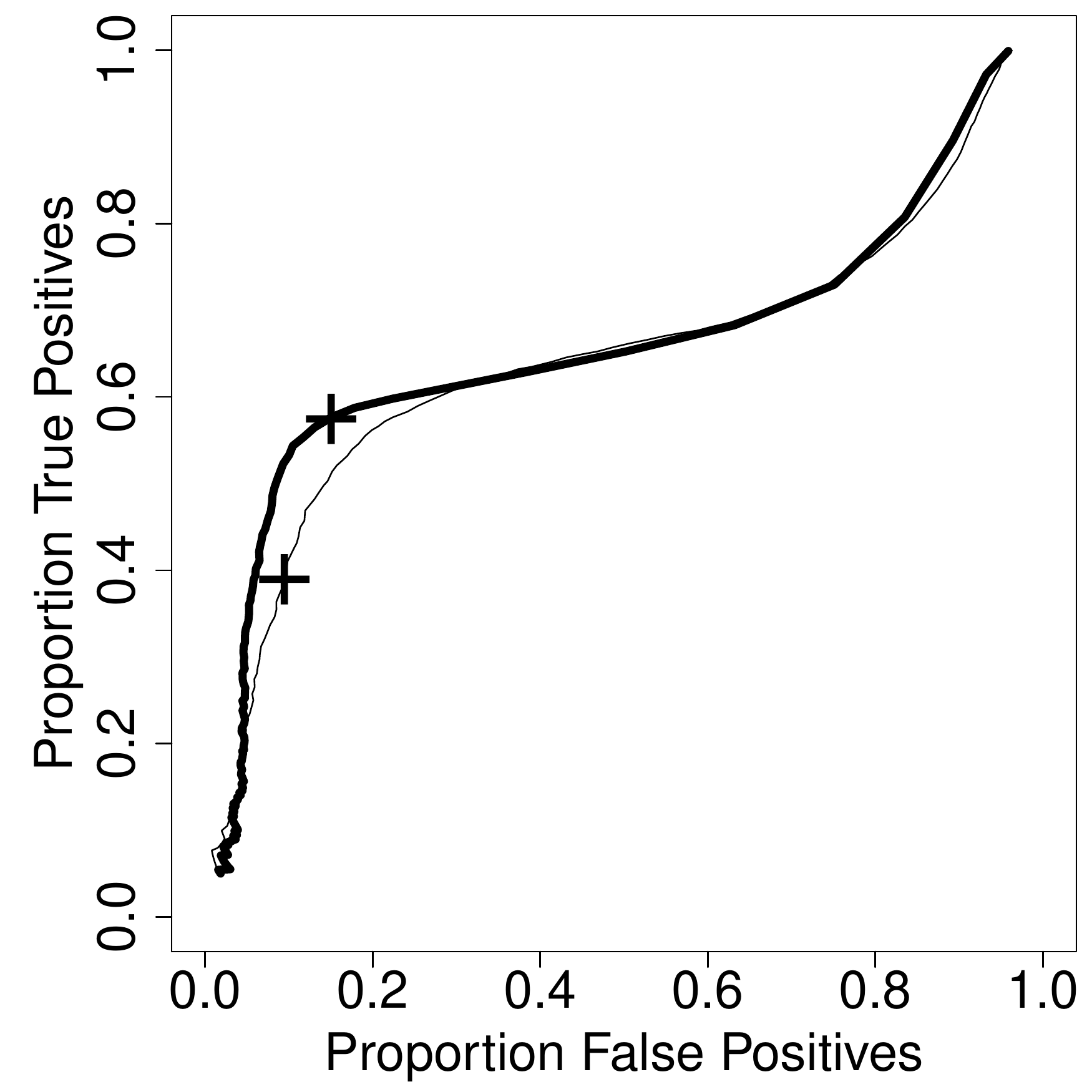}\label{fig:ftvar500}}
		\subfigure[]{\includegraphics[width=0.32\textwidth]{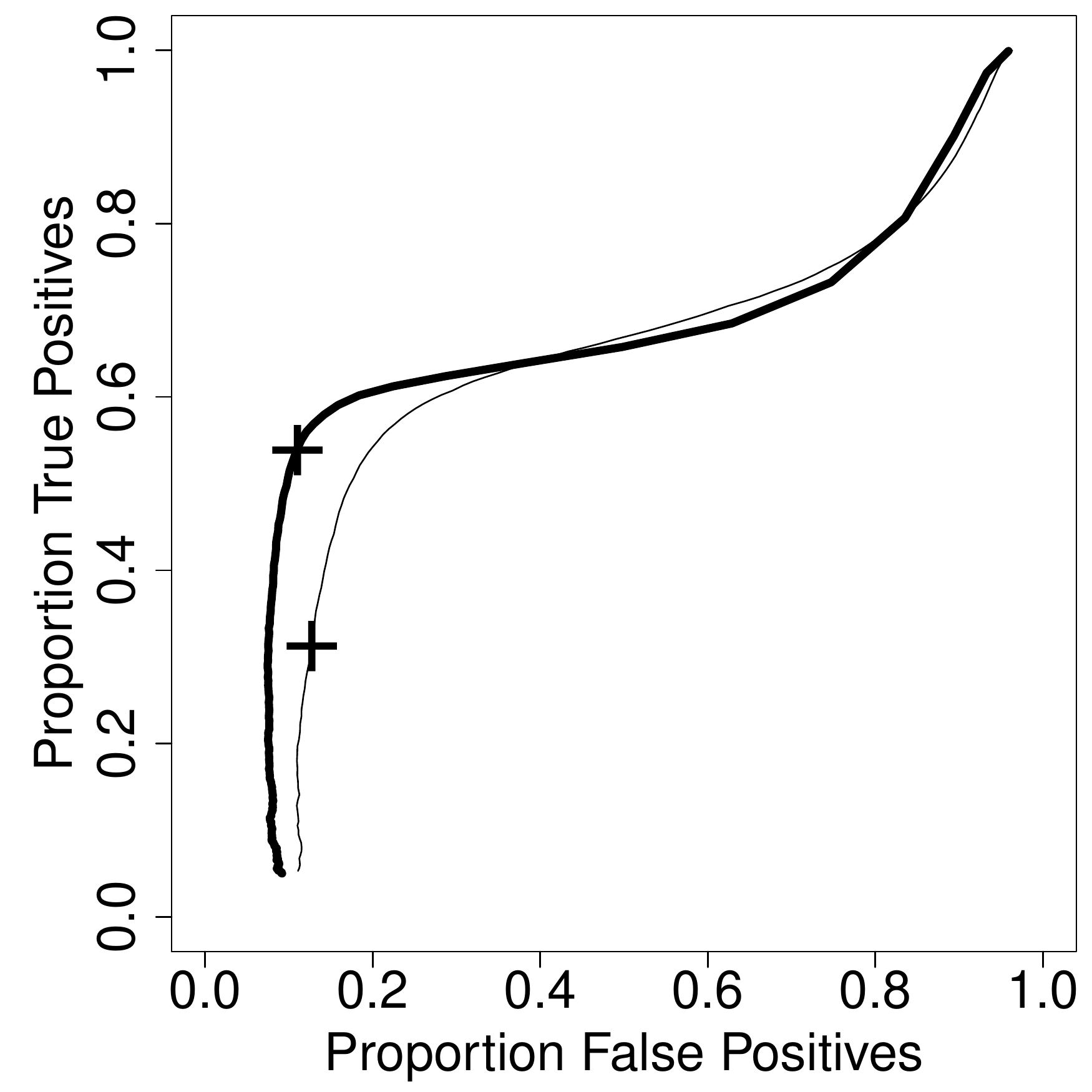}\label{fig:ftvar5000}}
		\subfigure[]{\includegraphics[width=0.32\textwidth]{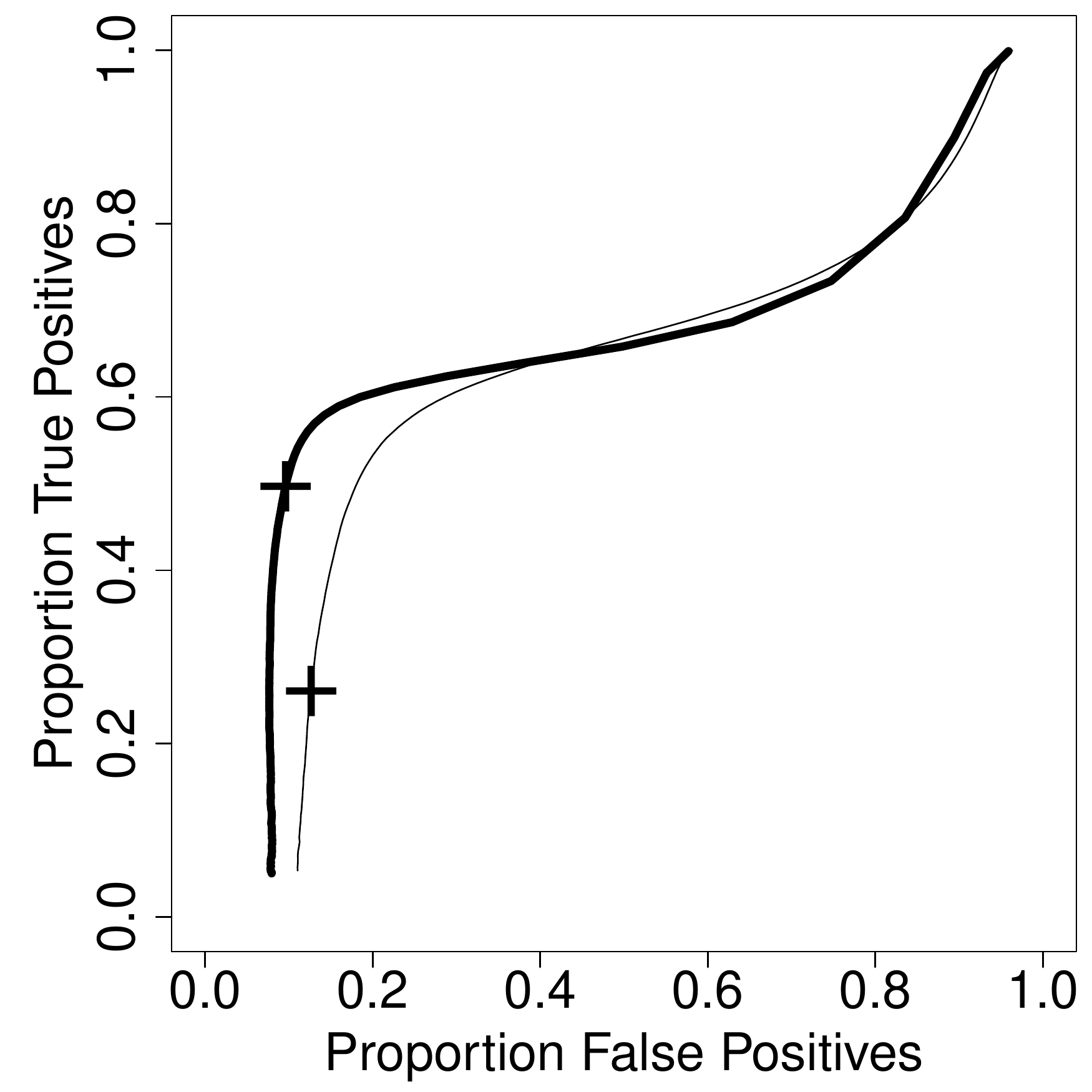}\label{fig:ftvar50000}}
\end{figure}

A final way to compare the accuracy of PELT with that of BS is to look at how accurately each method detects the actual times at which changepoints occurred. For the purposes of this study a
changepoint is considered correctly identified if we infer its location within a distance of 10 time-points of the true position. If two changepoints are identified in this window then one is counted
as correct and one false. The number of false changepoints is then the total number of changepoints identified minus the number correctly identified.  The results are depicted in Figure
\ref{fig:FTvar} for a selection of data lengths, $n$, for the case $m=n/100$.  As $n$ increases the difference between the PELT and BS algorithms becomes clearer with PELT correctly identifying more
changepoints than BS.  
Qualitatively similar results are obtained if we change how close an inferred changepoint has to be to a true changepoint to be classified as correct. Figures for square root increasing and fixed
numbers of changepoints are given in the supplementary material.  
As the number of changepoints decreases a higher proportion of true changepoints are detected with fewer false changepoints.

The supplementary material also contains an exploration of the same properties for changes in both mean and variance.  The results are broadly similar to those described above.  We now demonstrate increased accuracy of the PELT algorithm compared with BS on an oceanographic data set; a financial application is given in the supplementary material.

\subsection{Application to Canadian Wave Heights}\label{sec:NS}
There is interest in characterising the ocean environment, particularly in areas where there are marine structures, e.g. offshore wind farms or oil installations. Short-term operations, such as
inspection and maintenance of these marine structures, are typically performed in periods where the sea is less volatile to minimize risk.

\begin{figure}[t]
	\caption{{\footnotesize North Atlantic Wave Heights (a) Original data (b) Differenced data with PELT changepoints (c) Differenced data with optimal BS changepoints and additional subBS changepoints
(dotted lines).}}\label{fig:Canada}
	\centering
		\subfigure[]{\includegraphics[width=\textwidth]{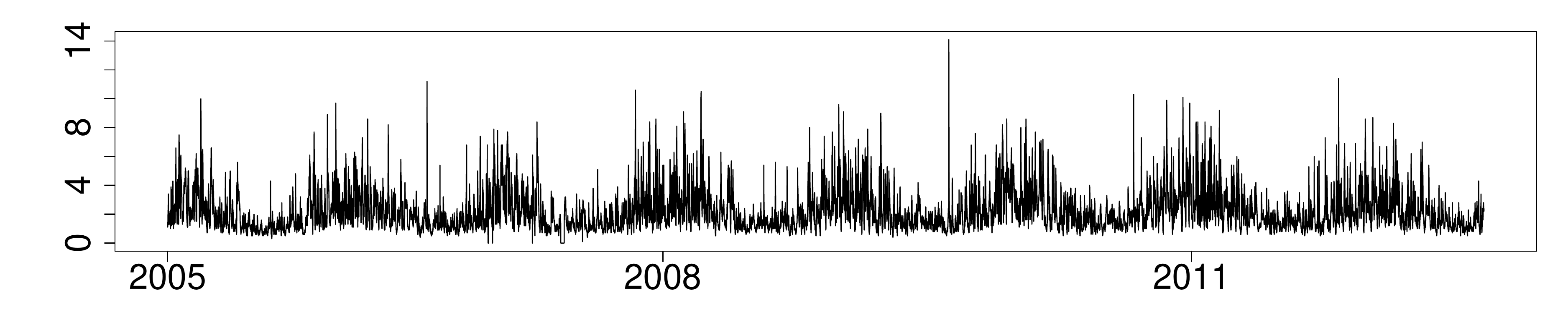}\label{fig:CanadaData}}
		\subfigure[]{\includegraphics[width=\textwidth]{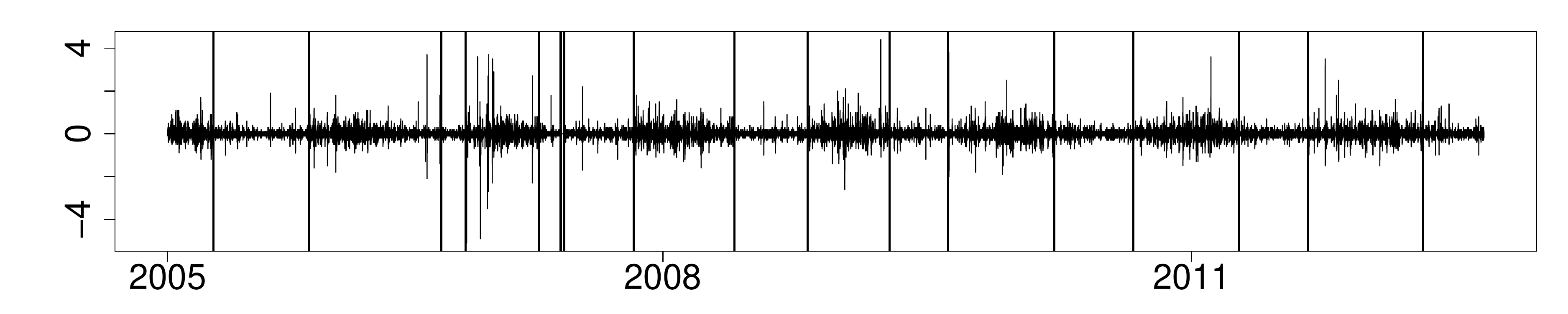}\label{fig:dCanadaPELT}}
		\subfigure[]{\includegraphics[width=\textwidth]{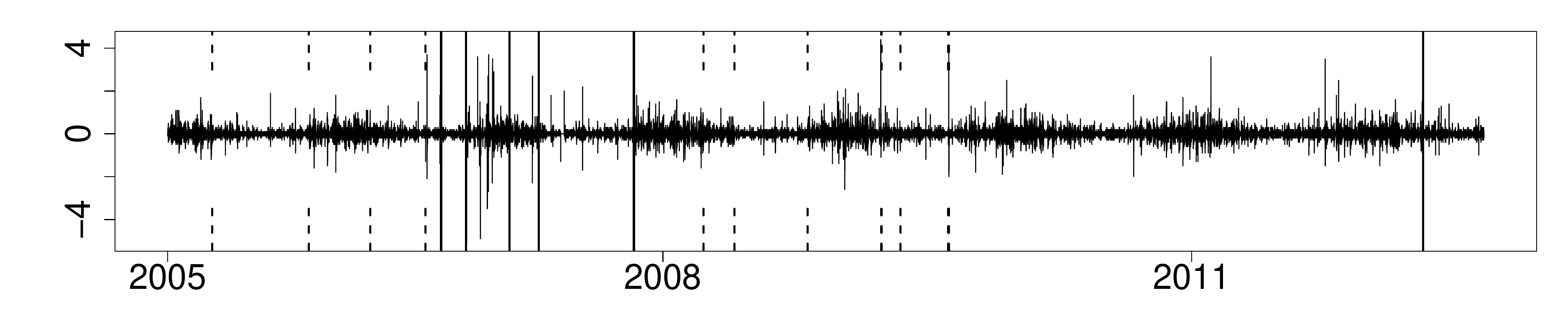}\label{fig:dCanadaBS}}
\end{figure}

Here we consider publically available data for a location in the North Atlantic where data has been collected on wave heights at hourly intervals from January 2005 until September 2012, see Figure
\ref{fig:CanadaData}. Our interest is in segmenting the series into period of lower and higher volatility. The data we use is obtained from Fisheries and Oceans Canada, East Scotian Slop buoy ID
C44137 and has been reproduced in the \texttt{changepoint} R package \citep{changepoint}.

The cyclic nature of larger wave heights in the winter and small wave heights in the summer is clear. However, the transition point from periods of higher volatility (winter storms) to lower
volatility (summer calm) is unclear, particularly in some years. To identify these features we work with the first difference data. Consequently a natural approach is to use the change in variance
cost function of Section \ref{Sec:changevar}. Of course this is but one of several ways in which the data could be segmented.

For the data we consider (Figure \ref{fig:CanadaData}) there is quite a difference in the number of changepoints identified by PELT (17) and optimal Binary Segmentation (6). However, the location of
the detected changepoints is quite similar. The difference in likelihood between the inferred segmentations is 3851. PELT chooses a segmentation which, by-eye, segments the series well into the
different volatility regions (Figure \ref{fig:dCanadaPELT}). Conversely, the segmentation produced by BS does not (Figure \ref{fig:dCanadaBS}); most notably it fails to detect any transitions between
2008 and 2012. If we increase the number of changepoints BS finds to equal that of PELT, the additional changepoints still fail to capture the regions appropriately.

\subsection{Changes in Auto-covariance within Autoregressive Data} \label{sec:AR}
Changes in AR models have been considered by many authors including \cite{DavisLeeRodriguezYam2006}, \cite{HuskovaPraskovaSteineback2007} and \cite{Gombay2008}.  
This section describes a simulation study that compares the properties of PELT and the genetic algorithm used in \cite{DavisLeeRodriguezYam2006} to implement the minimum description length (MDL) test statistic.  

\paragraph{{\itshape Minimum Description Length for AR Models}}
The simulation study here will be constructed in a similar way to that of Section \ref{Sec:varsimstudy}. 
 It is assumed that the data follow an autoregressive model with order and parameter values depending on the segment.  We shall take the cost function to be the MDL, 
and consider allowing AR models of order $1,\ldots,p_{max}$, for some chosen $p_{max}$ within each segment. The associated cost for a segment is
 \begin{align}
 	\mathcal{C}(y_{(\tau_{i-1}+1):\tau_i}) = \min_{p\in\{1,\ldots,p_{max}\}}\left\{ \log{p} + \frac{p+2}{2}\log(\tau_i-\tau_{i-1}) + \frac{\tau_i-\tau_{i-1}}{2}\log{\left(2\pi\hat{\sigma}(p,\tau_{i-1}+1,\tau_i)^2\right)} 
\right\}.
 \end{align}
where $\hat{\sigma}(p,\tau_{i-1}+1,\tau_i)^2$ is the Yule-Walker estimate of the innovation variance for data $y_{(\tau_{i-1}+1):\tau_i}$ and order $p$.  
When implementing PELT, we set $K=-[2\log(p_{max})+(p_{max}/2)\log(n)]$, to ensure that \eqref{ass:cpt} is satisfied.

\paragraph{{\itshape Simulation Study}}
The simulated data consists of 5 scenarios with varying lengths, $n=c$(1000, 2000, 5000, 10000, 20000) and each scenario contains $0.003n$ changepoints. 
 These changepoints are distributed uniformly across $(2,n-2)$ with the constraint that there must be at least 50 observations between changepoints. 
 Within each of these 5 scenarios we have 200 repetitions where the segment order is selected randomly from $\{0,1,2,3\}$ and the autoregressive parameters for each segment are a realisation from a 
standard Normal distribution subject to stationarity conditions.  We compare the output from PELT with an approximate method proposed by \cite{DavisLeeRodriguezYam2006} for minimising the MDL criteria, which uses a genetic algorithm. 
This was implemented in the program Auto-PARM, made available by the authors. We used the recommended settings except that for both methods we assumed $p_{max}=7$.  

 Table \ref{tbl:MDL} shows the average difference in MDL over each scenario for each fitted model.  It is clear that on average PELT achieves a lower MDL than the Auto-PARM algorithm and that this 
difference increases as the length of the data increases. Overall, for 91\% of data sets, PELT gave a lower value of MDL than Auto-Parm.  In addition, the average number of iterations required for PELT to converge is small in all cases. 

\begin{table}[t]
	\caption{\label{tbl:MDL}Average MDL and number of PELT iterations over 200 repetitions.}
	\centering
	\begin{tabular}{l||ccccc}\hline
		$n$ & 1,000 & 2,000 & 5,000 & 10,000 & 20,000 \\
		no. iterations & 2.470 & \ 2.710 & \ 2.885 & \ \ 2.970 & \ \ 3.000 \\
		Auto-PARM - PELT & 8.856 & 13.918 & 59.825 & 252.796 & 900.869 \\ \hline
	\end{tabular}
\end{table}

Previously, it was noted that the PELT algorithm for the MDL penalty is no longer an exact search algorithm.  
For $n=1,000$ we evaluated the accuracy of PELT by calculating the optimal segmentation in each case using Segment Neighbourhood (SN).
The average difference in MDL between the SN and PELT algorithms is $1.01$ (to 2dp). However SN took an order of magnitude longer to run than PELT, its computational cost increasing with the cube of the data size making it impracticable for large $n$. A better approach to improve on the results of our analysis would be to improve the search strategy 
for the value of penalty function to run PELT with.

\section{DISCUSSION}\label{Sec:conc}
In this paper we have presented the PELT method; an alternative exact multiple changepoint method that is both computationally efficient and versatile in its application. 
 It has been shown that under certain conditions, most importantly that the number of changepoints is increasing linearly with $n$, the computational efficiency of PELT is $\mathcal{O}(n)$. 
The simulation study and real data examples demonstrate that the assumptions and conditions are not restrictive and a wide class of cost functions can be implemented. 
The empirical results show a resulting computational cost for PELT that can be orders of magnitude smaller than alternative exact search methods. Furthermore, the results show 
substantial increases in accuracy by using PELT compared with Binary Segmentation.  Whilst PELT is not, in practice, computationally quicker than Binary Segmentation, 
we would argue that the statistical benefits of an exact segmentation outweigh the relatively small computational costs. There are other fast algorithms for segementing data that
improve upon Binary Segmentation \citep{GeyLebarbier2008,HarchaouiLevyLeduc2010}, although these do not have the guarantee of exactness that PELT does.

\cite{Rigaill2010} develops a competing exact method called pruned dynamic programming (PDPA). This method also aims to improve the computational efficiency of an exact method, this time Segment
Neighbourhood, through pruning, but the way pruning is implemented is very different from PELT. The methods are complementary. Firstly they can be applied to different problems, with PDPA able to cope
with a non-linear penalty functions for the number of changepoints, but restricted to models with a single parameter within each segment. Secondly the applications under which they are computationally
efficient is different, with PDPA best suited to applications with few changepoints. Whilst unable to compare PELT with PDPA on the change in variance or the change in mean and variance models
considered in the results section, we have done a comparison between them on a change in mean. Results are presented in Table 1 of the Supplementary material. Our comparison was for both a linearly
increasing number of changepoints, 
and a 
fixed number of changepoints scenario. For the former PELT was substantially quicker, by a factors of between 300 and 40,000 as the number of data-points varied between 500 and 500,000. When we fixed the number of changepoints to 2, PDPA was a factor of 2 quicker for data with 500,000 changepoints, though often much slower for smaller data sets.

Code implementing PELT is contained within the \texttt{R} library \texttt{changepoint} which is available on CRAN \citep{changepoint}.

\bibliography{BibTeXoutput}
\end{document}